\documentclass[aps,prl,reprint,superscriptaddress]{revtex4-1}
\usepackage{todonotes}
\usepackage{graphicx,xcolor}
\usepackage{amsmath}
\usepackage{braket}
\usepackage{url}
\usepackage{filecontents}

\usepackage[pdftex,colorlinks=true,hypertexnames=false]{hyperref}

\newcommand{\HTO}{Ho$_2$Ti$_2$O$_7$}
\newcommand{\AIAO}{all-in--all-out}
\newcommand{\TITO}{two-in--two-out}
\newcommand{\TIOO}{three-in--one-out~/~one-in--three-out}

\newcommand{\rateHLD}{0.1~ms/T}
\newcommand{\rateToulouse}{25~ns/T}

\newcommand{\etal}{\textit{et al.} }
\newcommand{\kB}{\ensuremath{k_\mathrm{B}} }

\newcommand{\muB}{\ensuremath{\mu_\text{B}} }

\begin{document}


\title{Magnetization beyond the Ising limit of \HTO}

\author{L. Opherden} \email[]{l.opherden@hzdr.de}
 \affiliation{Hochfeld-Magnetlabor Dresden (HLD-EMFL), Helmholtz-Zentrum  Dresden-Rossendorf, 01328 Dresden, Germany}
 \affiliation{Institut f\"{u}r Festk\"{o}rper- und Materialphysik, TU Dresden, 01062 Dresden, Germany}

\author{T. Herrmannsd\"orfer}
\author{M. Uhlarz} 
\author{D.~I. Gorbunov} 
 \affiliation{Hochfeld-Magnetlabor Dresden (HLD-EMFL), Helmholtz-Zentrum Dresden-Rossendorf, 01328 Dresden, Germany}

\author{A. Miyata}
\author{O. Portugall}
   \affiliation{Laboratoire National des Champs Magnetiques Intenses (LNCMI-EMFL), CNRS-UJF-UPS-INSA, 143 Avenue de Rangueil, 31400 Toulouse, France.}
 
\author{I. Ishii}
  \affiliation{Department of Quantum Matter, AdSM, Hiroshima University, Higashi Hiroshima 739-8530, Japan}
  
\author{T. Suzuki}
 \affiliation{Department of Quantum Matter, AdSM, Hiroshima University, Higashi Hiroshima 739-8530, Japan}

\author{H. Kaneko}
\author{H. Suzuki}
\affiliation{Department of Physics, Kanazawa University, Kakuma-machi, Kanazawa 920-1192, Japan}
 	
\author{J. Wosnitza}
 \affiliation{Hochfeld-Magnetlabor Dresden (HLD-EMFL), Helmholtz-Zentrum Dresden-Rossendorf, 01328 Dresden, Germany}
 \affiliation{Institut f\"{u}r Festk\"{o}rper- und Materialphysik, TU Dresden, 01062 Dresden, Germany}

\date{\today}

\begin{abstract}%
We report that the local Ising anisotropy in pyrochlore oxides -- the crucial requirement for realizing the spin-ice state -- can be broken by means of high magnetic fields.
For the case of the well-established classical spin-ice compound \HTO\ the magnetization exceeds the angle-dependent saturation value of the Ising limit using ultra-high fields up to 120~T.
However, even under such extreme magnetic fields full saturation cannot be achieved.
Crystal-electric-field calculations reveal that a level crossing for two of the four ion positions leads to magnetization steps at 55 and 100 T.
In addition, we show that by using a field sweep rate in the range of the spin-relaxation time the dynamics of the spin system can be probed. 
Exclusively at \rateToulouse\ a new peak of the susceptibility appears around 2~T. We argue, this signals the cross-over between spin-ice and polarized correlations.
\end{abstract}


\maketitle

Cubic pyrochlore-oxides, $R_2T_2$O$_7$, with $R$ being typically a rare earth and $T$ a transition-metal element, are arguably one of the most intensively studied material classes of the last two decades in the context of frustrated magnetism.
Both, the $R^{3+}$  and the $T^{4+}$ ions, form a sublattice of corner-sharing tetrahedra.
A strong crystal electrical field (CEF), created by the surrounding oxygen ions, can lead to a pronounced local anisotropy of the spins.
Spin-spin interactions, such as dipole-dipole or super-exchange interaction, lead to a variety of novel states of matter.
In case of a local Ising anisotropy, such materials can adopt a spin-liquid state where no long-range order can be found down to lowest temperatures.
Instead, short-range correlations of the spins lead to emergent fractionalized excitations.

A famous example for such a novel state of matter is the classical spin-liquid state classified as \textquoteleft spin ice\textquoteright \cite{Balents_2010}.
The local ice-rules lead to an arrangement where two spins point into and two spins point out of each tetrahedron (\TITO) which leads to a macroscopic ground-state degeneracy as realized by a remaining zero-point entropy \cite{Ramirez_1999}.
This state of matter became especially topical by realizing that excitations out of the \TITO\ arrangement can be interpreted as a pair of magnetic monopoles which can deconfine with finite energy cost \cite{Castelnovo_2008, Gingras_2009, Fennell_2009}.

The classical spin-ice state is realized for the six dysprosium and holmium-based pyrochlores Ho$_2T_2$O$_7$ and Dy$_2T_2$O$_7$ ($T$ = Ti, Ge, Sn) \cite{Zhou_2012}.
For a different ratio of exchange to dipole interaction \cite{denHertog_2000}, Ising spins on this lattice may order antiferromagnetically in the \AIAO\ state for which recently the exotic phenomenon of an inverse hysteresis was discovered in Nd$_2T_2$O$_7$ ($T$ = Zr, Hf) \cite{Opherden_2017, Opherden_2018}.
For the praseodymium-based pyrochlores Pr$_2T_2$O$_7$ ($T$ = Zr, Hf, Sn) \cite{Lee_2012, Kimura_2013, Sibille_2016-1, Zhou_2008, Anand_2016} a quantum spin-ice state is reported to be realized.
For all of these phenomena a local Ising anisotropy is crucial.
However, not much is known about the stability of this fundamental requirement.

Here, we probe the robustness of this local Ising anisotropy in ultra-strong magnetic fields using the classical spin-ice compound, \HTO.
Magnetization measurements were performed using high non-destructive magnetic-field pulses up to 58~T at the Dresden High Magnetic Field Laboratory as well as ultra-high semi-destructive pulses \cite{Portugall_1999} up to 130~T at the Laboratoire National des Champs Magnetiques Intenses at temperatures significantly higher than the spin-spin interaction energies.
In this case the magnetization can be treated by using a single-ion model.
The magnetization was measured using a compensated pair of coils as described elsewhere \cite{Skourski_2011, Takeyama_2011}.
For the non-destructive pulses the field changed by 55~T in 5.3~ms leading to a field sweep rate of about \rateHLD.
In the case of the semi-destructive experiments, the same magnetic-field range was reached in only 1.4~$\mu$s corresponding to a 4000 times faster field sweep rate of about \rateToulouse\ \footnote{For non-destructive pulses only up-sweep data is shown due to a lower level of noise. However, no qualitative difference was observed between up- and down-sweep data. For semi-destructive pulses down sweeps are discussed because up sweeps are strongly affected by discharge noise coming from spark gap switches.}.
Experimental results are compared to CEF calculations and to static magnetization measurements up to 14~T which were obtained by using a commercial vibrating-sample magnetometer.
All measurements in pulsed fields were performed along the long axis of the used crystal in order to obtain a good filling factor of the pick-up coil. This corresponds to the orientation $H\|$[5$\,$5$\,$13] (see Fig. \ref{Fig5}). Samples were grown by the floatingzone method as described in \cite{Fukazawa_2002}.

\begin{figure*}[t]
	\includegraphics[width=.8\linewidth]{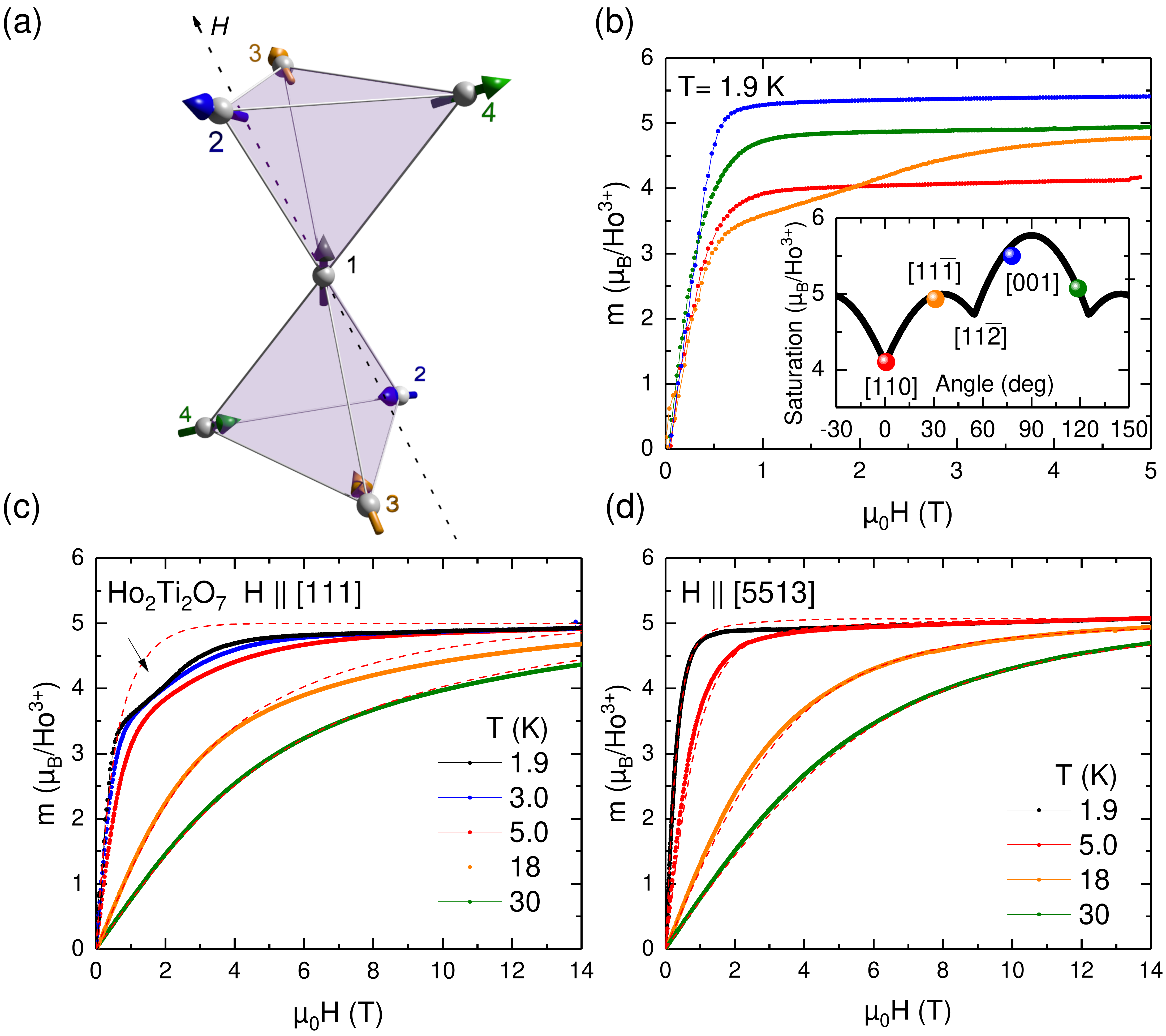}
	\caption{(a) Illustration of two tetrahedra and their four local Ising directions. The dashed line indicates the used direction of the magnetic field, $H\|$[5$\,$5$\,$13]. (b) Magnetization of \HTO\ for different orientations at 1.9 K. Inset: The achieved saturation value within the Ising limit is plotted against the calculated angular-dependent saturation magnetization using Eq. (\ref{eq:Magnetisierung_Tetraeder_Ising}) for a rotation perpendicular to [1$\bar{1}$0]. Displayed crystal directions label extremal points. The green and the blue dot corresponds to the direction [5$\,$5$\,$13] and [11$\bar{6}$] respectively. (c) Magnetization for different temperatures for a field parallel to [111]. The arrow indicates the kagome-ice plateau as a signature of the spin-ice phase.  Red dashed lines indicate the behavior within the Ising limit for 3, 18 and 30 K. (d) Magnetization for the orientation used in the pulsed-field experiment. Red dashed lines show the expected calculated behavior within the Ising limit.
		\label{Fig1}}
\end{figure*}
In case of $R_2T_2$O$_7$ spin-ice compounds, the $R$ ions contribute localized magnetic moments while the $T$ ions are non-magnetic.
Each $R$-ion is surrounded by eight non-equidistant oxygen ions which create a strong trigonal CEF \cite{Gardner_2010, Blanchard_2013}. The two closest O$^{2-}$ ions are arranged along the line connecting the centers of the tedrahedra making this the quantization or local $z$ axis for the CEF Hamiltonian \cite{Rosenkranz_2000},
\begin{equation}
\begin{split}
\hat{H}_\mathrm{CEF} &= B_0^2 \hat{C}_0^2 + B_0^4 \hat{C}_0^4 + B_3^4 (\hat{C}_3^4 - \hat{C}_{-3}^4) \\
&+ B_0^6 \hat{C}_0^6  + B_3^6 (\hat{C}_3^6-\hat{C}_{-3}^{6}) + B_6^6 (\hat{C}_6^6 + \hat{C}_{-6}^6) \, ,
\end{split}
\label{eq:CEF_Hamiltonian}
\end{equation}
where $B^k_q$ are the CEF parameters and $\hat{C}^k_q$ are tensor operators.
The strong CEF leads to a ground-state doublet of an almost pure $\ket{\pm} = \ket{\pm J}$ state and a first excited state at 230~K \cite{Ruminy_2016} in the case of \HTO.
Therefore, for low temperatures and low fields the system can be described as a pseudo spin-half state having an effective moment of $\mu = g_J J = $ 10 $\mu_\mathrm{B}$.
Because for such a state the magnetization, $m_i \propto \braket{\pm|\hat{J}_i|\pm}$, vanishes for the $x$ and $y$ component, the moments show a strong Ising anisotropy along their local quantization axis.
The four local Ising directions correspond to the four equivalent $\langle 111 \rangle$ directions of the cubic lattice [Fig. \ref{Fig1}(a)].
\begin{figure}[b]
	\includegraphics[width=0.7\linewidth]{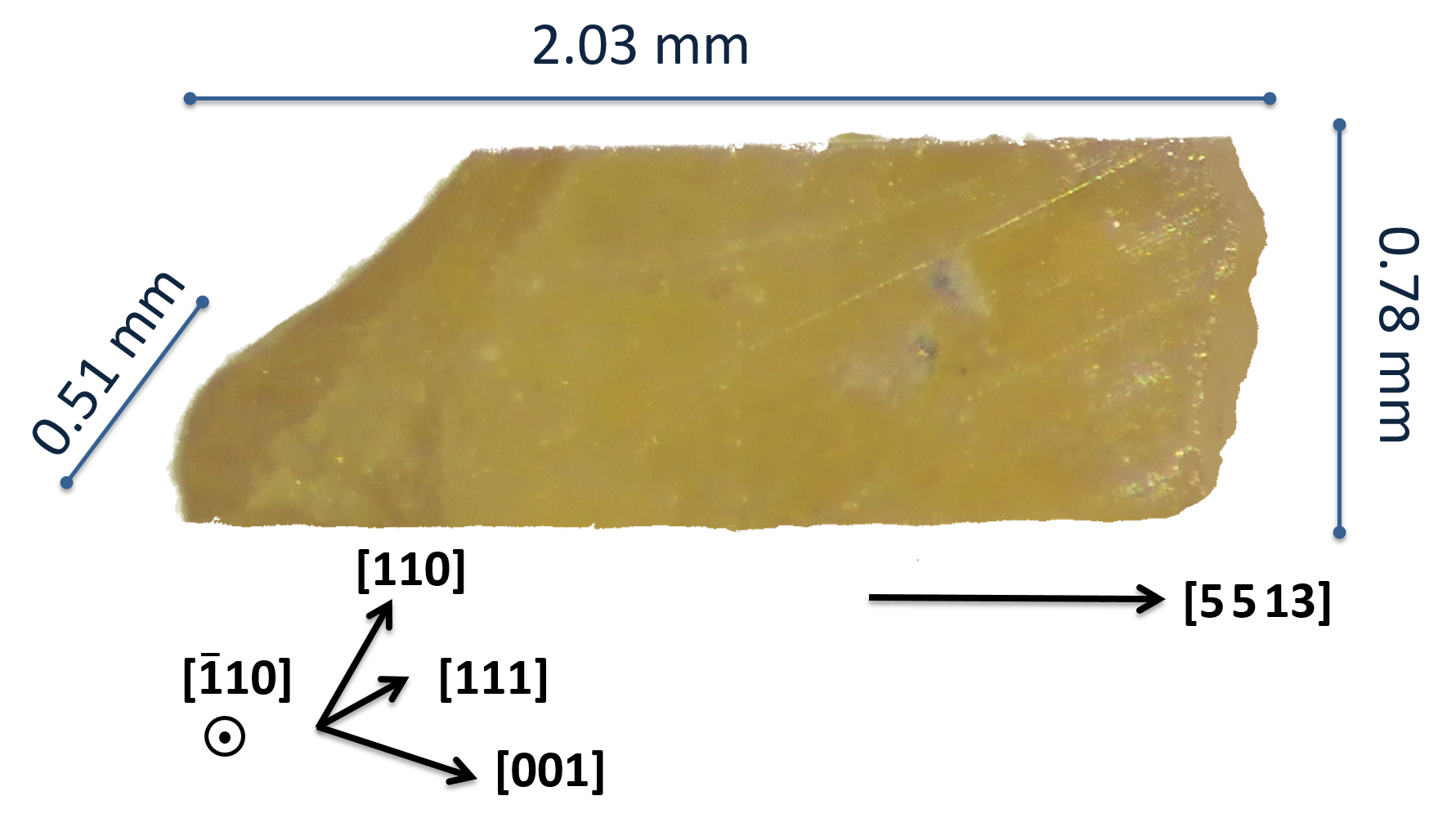}
	\caption{Photography of the used \HTO\ single crystal. The long side of the crystal is oriented along [5$\,$5$\,$13].
		\label{Fig5}}
\end{figure}
\begin{figure*}[t]
	\includegraphics[width=\linewidth]{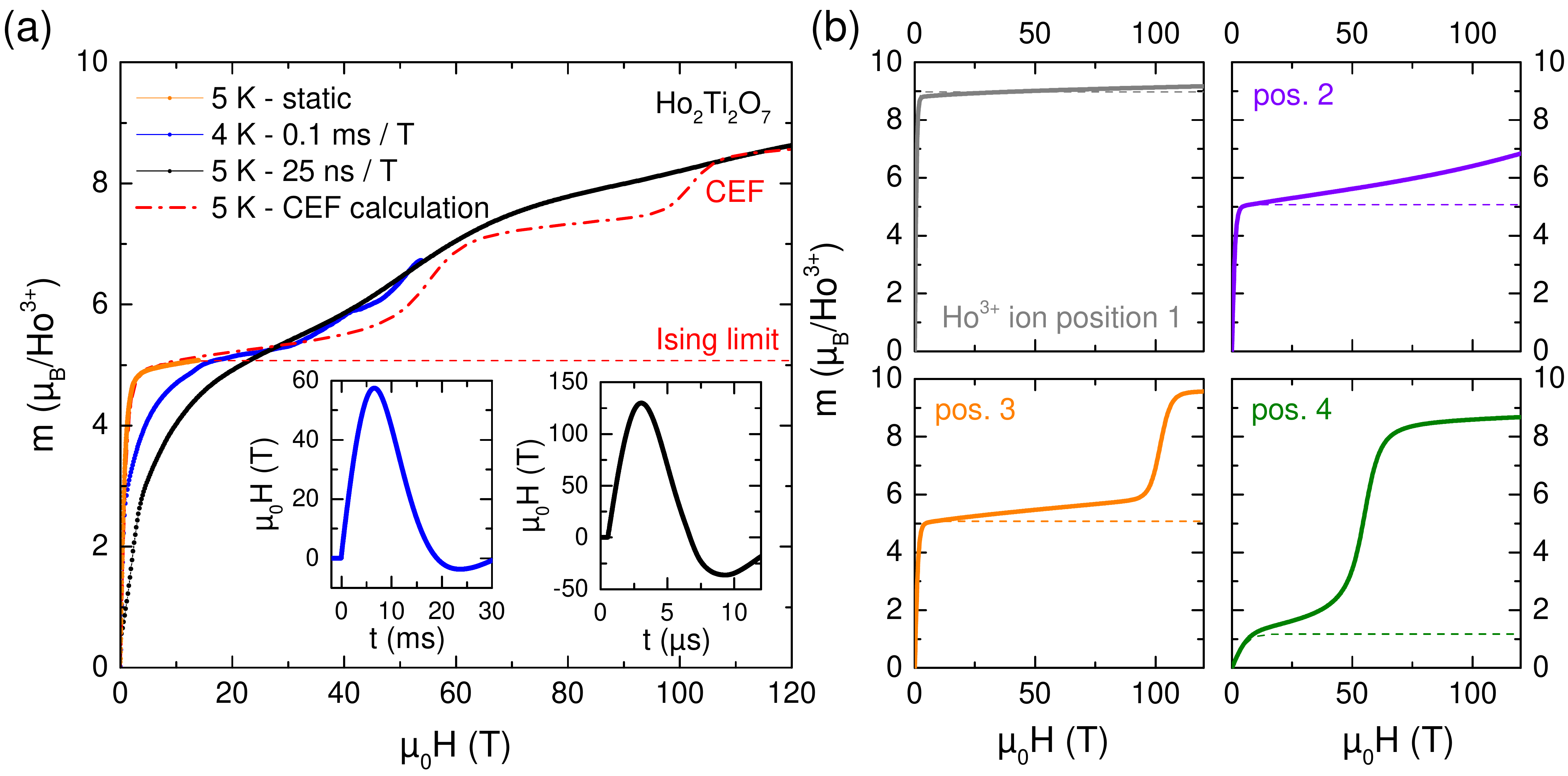}
	\caption{Field-dependent magnetization of \HTO\ for measurements in static fields (orange) and in pulsed fields in the ms (blue) and $\mu$s range (black) at 4 and 5 K for $H\|$[5$\,$5$\,$13].
	The red dashed line shows the calculated behavior of the Ising limit ny use of Eg. (\ref{eq:Magnetisierung_Tetraeder_Ising}), the dashed-dotted red line shows the result of the CEF calculations. Insets: Time dependence of magnetic field for the corresponding field pulses. (b) Calculated magnetization for the four individual ion positions illustrated in Fig. \ref{Fig1} (a).	
		\label{Fig2}}
\end{figure*}
Accordingly for non-interacting pseudo spin-half moments with perfect Ising anisotropy the magnetization can be described by 
\begin{equation}
M = \dfrac{1}{4} \sum_{i=1}^{4} \mu |\vec{n}_i \cdot \vec{h}| \tanh \left( \frac{\mu |\vec{n}_i \cdot \vec{h}| B}{\kB T} \right) \, , 
\label{eq:Magnetisierung_Tetraeder_Ising}
\end{equation}
where $\vec{n}_i$ are the normalized vectors of the $\langle 111 \rangle$ directions, $\vec{h} = \vec{H}/|\vec{H}|$ is the direction of the field, and $B = \mu_0 |\vec{H}|$.
Due to the different projections of the external magnetic field onto the four spin directions angle-dependent saturation values of the magnetization are reached in the Ising limit with the easy axis along the $\langle 001 \rangle$ and the hard axis along the $\langle 110 \rangle$ directions [Fig. \ref{Fig1}(b) with inset].
Eq. (\ref{eq:Magnetisierung_Tetraeder_Ising}) can describe the field-dependent magnetization of \HTO\ which was measured for static magnetic fields up to 14~T [red dashed lines in Fig. \ref{Fig1} (c,d)].
For a magnetic field applied along the [111] direction, a value of 5.0 \muB is achieved for the Ising-limit plateau [Fig. \ref{Fig1}(c)].
For a field oriented along [5$\,$5$\,$13], the alignment used as well in the pulsed-field experiments, a clear magnetization plateau appears at around 5.1 $\mu_\mathrm{B}$.
Both values show conformity with the theoretically expectation [high-field limit of Eq. (\ref{eq:Magnetisierung_Tetraeder_Ising})].
Whereas for $H\|$[5$\,$5$\,$13] the magnetic-field dependence of the magnetization can be well accounted for, deviations can be seen for $H\|$[111].
For the lowest temperatures an additional rounded plateau at $m$ = 3.3 \muB [arrow in Fig. \ref{Fig1}(c)] signals the direct evidence of the spin-ice phase which is not captured by the model of non-interacting Ising spins [red dashed lines in Fig. \ref{Fig1}(c)].
By applying a magnetic field, those \TITO\ configurations which have collinear spins parallel to the field are selected from the ground-state manifold.
This intermediate state is called kagome-ice phase.
At a field of about 2~T, the short-range order is broken and the sample adopts the polarized \TIOO\ orientation \cite{Krey_2012, Petrenko_2003, Fennell_2007}.

\begin{figure*}[t]
	\includegraphics[width=\linewidth]{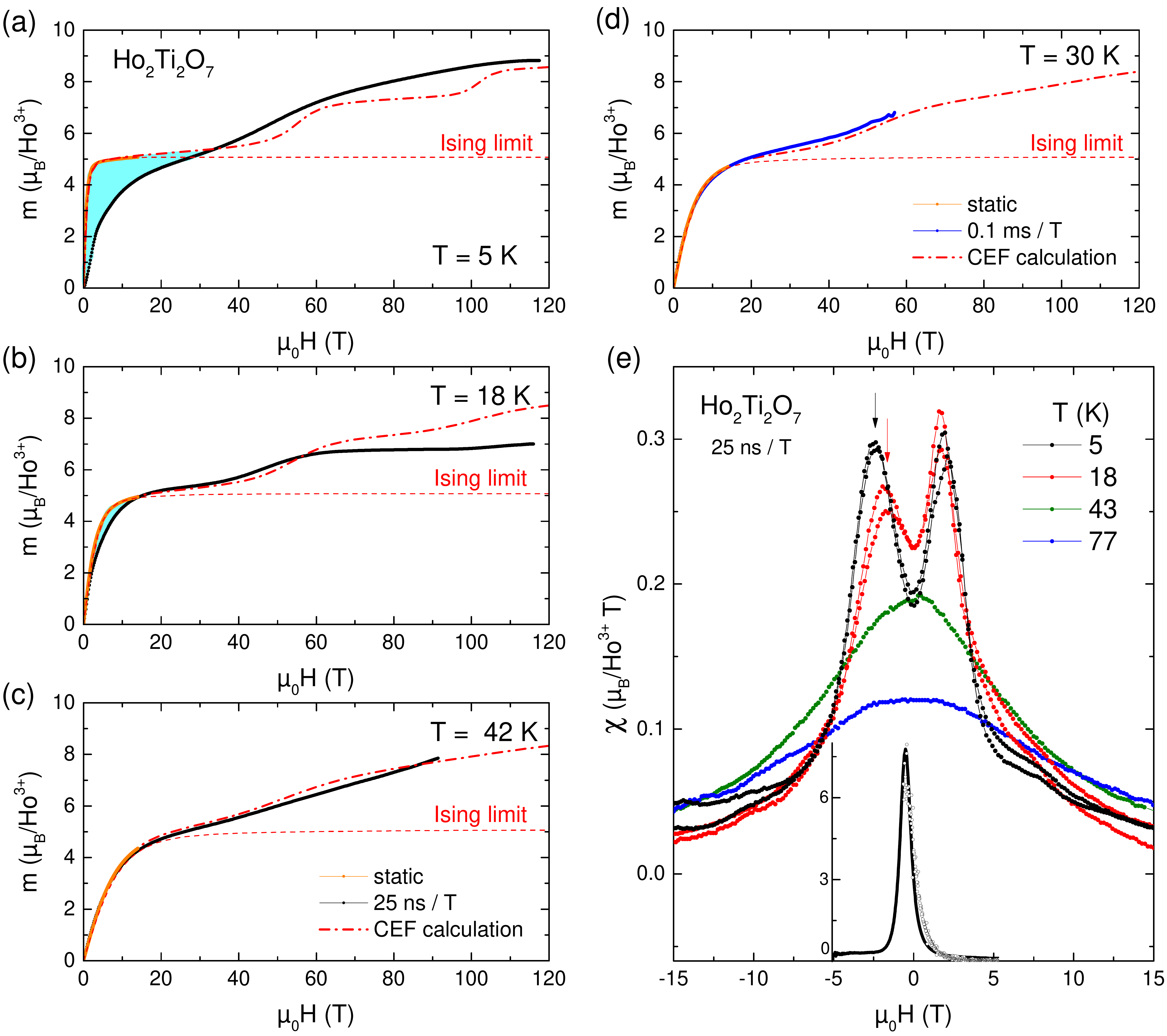}
	\caption{Field-dependent magnetization of \HTO\ of static field measurements (orange) and pulsed fields experiments in the $\mu$s range (black) at (a) 5 K, (b) 18 K respectively (c) 42 K and (d) for pulsed fields experiments in the ms (blue) at 30 K. All measured were performed in the orientation $H\|$[5$\,$5$\,$13]. (e) Field-dependent susceptibility for the pulsed-field measurements in the $\mu$s range  for $H\|$[5$\,$5$\,$13]. An additional peak appears in the data at 5 and 18 K. Inset: Feature-less behavior for static fields (open circles) and pulses in the ms-range (black line) at 4~K.
		\label{Fig3}}
\end{figure*}

Using pulsed magnetic fields [Fig. \ref{Fig2}], the local Ising anisotropy can be broken.
The magnetization clearly exceeds the Ising limit of 5.1 $\mu_\mathrm{B}$ at 15 and 25~T, respectively, depending on the field sweep rate as discussed below.
A step in the magnetization is seen at 55~T as predicted by CEF calculations.
A similar magnetic-field dependence was previously reported for $H\|$[111] \cite{Erfanifam_2014}.
However, even at 120~T full saturation,with all spins parallel to the field ($m_\mathrm{s}$ = 10 $\mu_\mathrm{B}$/Ho$^{3+}$), cannot be reached.
Interpolating the slope at highest fields, a saturation can be expected for fields not smaller than 180~T.

If an external magnetic field is applied, the single-ion Hamiltonian is given by
\begin{equation}
\hat{H} = \hat{H}_\mathrm{CEF} - g_J \mu_\mathrm{B}\hat{\vec{J}}\vec{H}  \, ,\\
\label{eq:CEF_Hamiltonian_im_Feld}
\end{equation}
where $\hat{H}_\mathrm{CEF}$ is given in Eq. (\ref{eq:CEF_Hamiltonian}) with parameters taken form Rosenkranz \etal \cite{Rosenkranz_2000}.
Because the magnetic field $\vec{H}$ needs to be transformed into the local coordinate system, the resulting eigenstates $\ket{n}$ and energies $E_i$ can differ for the four local positions of the holmium ions.
The expression for the used field direction in local coordinates can be found in the appendix.
The magnetization for each ion position as a function of the magnetic field is than given by \cite{VanHieu_2007, Kamikawa_2016}
\begin{equation}
M^\mathrm{Ion} =  \frac{\partial f}{\partial H} = \frac{g_J \mu_\mathrm{B}}{Z \hbar}\sum_{n} \braket{n|\hat{\vec{J}}|n}\exp \left( -\frac{E_n}{k_\mathrm{B}T} \right)  \ ,
\label{eq:M-Ion}
\end{equation}
with $Z = \sum_i \exp (-E_i/k_\mathrm{B}T)$ being the partition function and $f$ the free energy density.
The total magnetization of the spin system is obtained by the arithmetic mean of the four ion positions.
Note, that in the low-field limit Eq. (\ref{eq:Magnetisierung_Tetraeder_Ising}) is recovered.

By solving Eqs. (\ref{eq:CEF_Hamiltonian_im_Feld}) and (\ref{eq:M-Ion}) the magnetization was calculated for each ion position [Fig. \ref{Fig2}(b)].
A crossing of the energy levels at 55 and 100~T for the ion positions 4 and 3, respectively, leads to the noticeable steps in the calculated magnetization (Fig. \ref{Fig2}).
Because of the dynamic response of the magnetization, those steps are broadened in the experiment.
The step at 55~T can be seen clearly whereas the second step cannot be resolved.
For higher temperatures where static and pulsed-field magnetization data coincide, the CEF calculation fit the experimental data better [Figs. \ref{Fig3} (a-d)].

Recently, a vibronic bound-state was found to split up the $\Delta_5 = 60$ meV doublet of \HTO, when magneto-elastic interaction is considered additionally \cite{Ruminy_2016, Gaudet_2018-2}.
Due to the resolution of the magnetization measurements using $\mu$s pulses a clear signature of this phonon mediated process cannot be detected unambiguously.
However, it could explain deviations of the magnetization with respect to the CEF calculations at highest fields, where this doublets contributes most ($\Delta_5/\mu \approx$ 104 T).

The field sweep rate has a clear influence on the measured magnetization at low temperatures [Fig. \ref{Fig2}(a)].
Whereas for static measurements at 5~K the Ising plateau is reached at around 4~T, for the measurements in the ms range (\rateHLD) 15~T are necessary.
For magnetic pulses on a $\mu$s timescale even 25~T are needed. 
For the latter case, a considerable large field rate has been used, $\mathrm{d}\mu_0H / \mathrm{d}t$ = 1~T~/~25~ns.
The spin-relaxation time was determined to be $\sim$10~ns by means of inelastic neutron-scattering experiments at T = 1.4 K as well as 5 K in the zero-field limit  \cite{Clancy_2009}.
However, the field dependence of the magnetization in pulsed fields suggests distinctly longer spin-relaxation times in the range of milliseconds at this temperature.
This time scale follows the trend of low-temperature relaxation rates determined by various ac susceptibility measurements \cite{Quilliam_2011} and suggests a rather similar behavior compared to the relaxation rate of Dy$_2$Ti$_2$O$_7$ which is about 100 $\mu$s at 18~K \cite{Snyder_2003}.
We conclude, that the relaxation process mediating change of the magnetization is different compared to the one probed by the full-width at half-maximum of the backscattering energy scans observed by means of neutron scattering  (see Fig. 6 in ref. \cite{Clancy_2009}). 
A recent theoretical work predicts also two distinct relaxation times for spin-ice systems \cite{Tomasello_2018}.
However, both anticipated relaxation times should be considerable longer than the results from neutron scattering.

For higher temperatures the spin-relaxation time is decreasing as it can be seen by smaller difference between the pulse-field magnetization measurements and the static field results [light-blue shaded areas in Figs. \ref{Fig3} (a,b)].
Coincidence was found to be reached at 42~K in the $\mu$s range [Fig. \ref{Fig3} (c)] respectively at 30~K for experiments in the ms range [Fig. \ref{Fig3} (d)].

In addition, a change in the slope of the magnetization can be seen at low fields exclusively when the field-changing rate was in the order of $\mu$s.
The derivative of the magnetization data shows, therefore, a peak centered between 2 and 2.5~T [Fig. \ref{Fig3} (e)].
This matches the field where the kagome-ice state is broken for the orientation $H\|$[111].
Compared to the signatures of the kagome-ice phase the peak can be seen at distinct higher temperatures up to 18~K.
We presume, that this peak is the dynamic signature of the preferred correlation changing from \TITO\ to three-in--one-out. 
Such a transition might be associated with an increase of the spin-relaxation time seen as a change of slope of the magnetization when using fast field-sweep rates.

In summary, we demonstrate that the local Ising anisotropy of pyrochlore systems -- the key property for observing spin-ice and \AIAO \ behavior -- can be broken using high magnetic fields.
For the classical spin-ice compound \HTO\ the plateau of the Ising limit is exceeded at fields about around 25~T, however, full saturation cannot be achieved even at the ultra-high magnetic fields of 120~T in the presented orientation.
Using crystal-field calculations to determine the magnetization for each of the four ion positions, the field dependence of the magnetization can be described.
In particular, level crossings for two ion positions lead to magnetization steps at 55 and 100~T.
Whereas the first step can be resolved clearly, the second step is strongly suppressed in the experimental data.
An additional peak in the susceptibility, appearing exclusively at about 2~T for temperatures up to 18~K  if the field sweep rate is in the order of the spin-relaxation time, may be attributed to a cross-over for the preferred spin-arrangement changing from \TITO\ to the polarized configuration.
\\
\\
We acknowledge support by DFG through SFB 1143 and by HLD at HZDR and LNCMI-T, both members of the European Magnetic Field Laboratory (EMFL).
This work was also supported by JSPS KAKENHI Grant Number 17H06136 and CResCent (Chirality Research Center) in Hiroshima University (the MEXT program for promoting the enhancement of research universities, Japan). We thank T. Gottschall for providing the visualization.

\section{APPENDIX}
All measurements were performed in the orientation $H\|$[5$\,$5$\,$13].
To transform the global field direction $\vec{H}_\mathrm{global} = \frac{1}{\sqrt{219}}(5,5,13)^\mathrm{T}$ into the local coordinate system of the four ion positions, the field vector has to be multiplied with one of the following matrices, $\vec{H}_i = \hat{T}^i\vec{H}_\mathrm{global}$,\\

\begin{minipage}[t]{0.22\textwidth}
	\centering
	$\hat{T}^1$ = 
	$\begin{pmatrix}
	0 & \frac{2}{\sqrt{6}} & \frac{1}{\sqrt{3}} \\
	\frac{1}{\sqrt{2}} & -\frac{1}{\sqrt{6}} & \frac{1}{\sqrt{3}} \\
	-\frac{1}{\sqrt{2}} & -\frac{1}{\sqrt{6}} & \frac{1}{\sqrt{3}} \\
	\end{pmatrix}$
\end{minipage}
\begin{minipage}[t]{0.22\textwidth}
	\centering
	$\hat{T}^2$ = $\begin{pmatrix}
	0 & \frac{2}{\sqrt{6}} & -\frac{1}{\sqrt{3}} \\
	-\frac{1}{\sqrt{2}} & \frac{1}{\sqrt{6}} & \frac{1}{\sqrt{3}} \\
	\frac{1}{\sqrt{2}} & \frac{1}{\sqrt{6}} & \frac{1}{\sqrt{3}} \\
	\end{pmatrix}$
\end{minipage}

\begin{minipage}[t]{0.22\textwidth}
	\centering
	$\hat{T}^3$ = 
	$\begin{pmatrix}
	0 & -\frac{2}{\sqrt{6}} & \frac{1}{\sqrt{3}} \\
	\frac{1}{\sqrt{2}} & -\frac{1}{\sqrt{6}} & -\frac{1}{\sqrt{3}} \\
	\frac{1}{\sqrt{2}} & \frac{1}{\sqrt{6}} & \frac{1}{\sqrt{3}} \\
	\end{pmatrix}$
\end{minipage}
\begin{minipage}[t]{0.22\textwidth}
	\centering
	$\hat{T}^4$ = 
	$\begin{pmatrix}
	0 & \frac{2}{\sqrt{6}} & -\frac{1}{\sqrt{3}} \\
	\frac{1}{\sqrt{2}} & -\frac{1}{\sqrt{6}} & -\frac{1}{\sqrt{3}} \\
	\frac{1}{\sqrt{2}} & \frac{1}{\sqrt{6}} & \frac{1}{\sqrt{3}} \\
	\end{pmatrix}$
\end{minipage}
\\
which leads to
\begin{equation}
\begin{split}
\centering
\vec{H}_1 &= \left(\begin{array}{c} -4\sqrt{6} \\ -4\sqrt{2} \\ 23 \end{array}\right) \qquad
\vec{H}_2 = \left(\begin{array}{c} 4\sqrt{6} \\ 14\sqrt{2} \\ 13 \end{array}\right) \\
\vec{H}_3 &= \left(\begin{array}{c} 9\sqrt{6} \\  -1\sqrt{2} \\ 13 \end{array}\right) \qquad
\vec{H}_4 = \left(\begin{array}{c} 9\sqrt{6} \\ 9\sqrt{2} \\ 3 \end{array}\right) \, .
\label{eq:J-in-global-coords}
\end{split}
\end{equation}
\\
For the normalized directions each vector has to be multiplied by a normalization factor, $\vec{h}_i = \vec{H}_i / \left( 3\sqrt{73} \right)$.
Note, that the expected value of the Ising plateau stays unchanged supporting the validity of this result:
\begin{equation}
\begin{split}
m_\mathrm{max} &= \dfrac{\mu}{4} \sum_{i=1}^{4} |\vec{n}_i\cdot (5,5,13)^\mathrm{T}/\sqrt{219}| \\
&= \dfrac{\mu}{4} \sum_{i=1}^{4} |(0,0,1)^\mathrm{T}\cdot \vec{h}_i| \approx 5.07 \mu_\mathrm{B} \, ,
\end{split}
\end{equation} 
with $\mu$ = 10 $\mu_\mathrm{B}$.
The magnetization within the Ising limit is, thus, given by:
\begin{equation}
\begin{split}
m &= \frac{\mu}{3\sqrt{73}} \bigg[ 23 \tanh \left(\frac{23 \mu B}{3\sqrt{73} \kB T} \right)\\
 &+ 26 \tanh \left(\frac{13 \mu B}{3\sqrt{73} \kB T} \right) + \tanh \left(\frac{\mu B}{3\sqrt{73} \kB T} \right) \bigg] \, .
\end{split}
\label{eq:Magnetisierung_5513_Ising-Limit}
\end{equation}


%

\end{document}